\documentclass[draftclsnofoot, 12pt, onecolumn]{IEEEtran}

\usepackage{cite}
\usepackage{graphicx}
\usepackage{mathrsfs}
\usepackage{amssymb}
\usepackage{amsmath}
\usepackage{subeqnarray}
\usepackage{cases}
\usepackage{booktabs}
\usepackage{amsthm}
\usepackage{caption}
\usepackage{subfigure}
\usepackage{algorithm}
\usepackage{algorithmic}
\usepackage{url}
\usepackage{amsmath,amssymb}
\usepackage{graphicx}
\usepackage{url}
\usepackage{booktabs}
\usepackage{multirow}
\usepackage{subfigure}
\usepackage{authblk}
\usepackage{amsthm}
\usepackage{bm}
\usepackage{algorithm}
\usepackage{stfloats}
\usepackage{array}
\usepackage{eqparbox}
\usepackage{bm}
\usepackage{lipsum}
\usepackage{amsthm,amsmath,amssymb}
\usepackage{mathrsfs}
\usepackage{color}
\usepackage{makecell}
\usepackage{diagbox}
\allowdisplaybreaks[4]

\begin{document}



\title{Joint Optimization of Beam-Hopping Design and NOMA-Assisted Transmission for Flexible Satellite Systems}

\author{Anyue~Wang,
~\IEEEmembership{Graduate Student Member,~IEEE,}
        Lei~Lei,
~\IEEEmembership{Member,~IEEE,}
        Eva~Lagunas,
~\IEEEmembership{Senior Member,~IEEE,}
        Ana~I.~P\'{e}rez-Neira,
~\IEEEmembership{Fellow,~IEEE,}
        Symeon~Chatzinotas,
~\IEEEmembership{Senior Member,~IEEE,}
        and~Bj\"{o}rn~Ottersten,
~\IEEEmembership{Fellow,~IEEE}
\thanks{This work was supported in part by the Luxembourg National Research Fund (FNR) CORE projects ROSETTA (under grant 11632107) and FlexSAT (under grant 13696663). (Corresponding author: Lei Lei)}
\thanks{A part of this work \cite{wang2021joint} was presented at IEEE Wireless Communications and Networking Conference (WCNC), 2021.}
\thanks{A. Wang, L. Lei, E. Lagunas, S. Chatzinotas, and B. Ottersten are with Interdisciplinary Center for Security, Reliability and Trust, University of Luxembourg, 1855 Luxembourg (email: anyue.wang; lei.lei; eva.lagunas; symeon.chatzinotas; bjorn.ottersten@uni.lu).}
\thanks{A. I. P\'{e}rez-Neira is with the Centre Tecnol\`{o}gic de Telecomunicacions de Catalunya (CTTC/CERCA), 08860 Castelldefels, Spain, and Universitat Polit\`{e}cnica de Catalunya (UPC), 08034 Barcelona, Spain (email: ana.perez@cttc.es).}}

\maketitle

\begin{abstract} 
Next-generation satellite systems require more flexibility in resource management such that available radio resources can be dynamically allocated to meet time-varying and non-uniform traffic demands. 
Considering potential benefits of beam hopping (BH) and non-orthogonal multiple access (NOMA), we exploit the time-domain flexibility in multi-beam satellite systems by optimizing BH design, and enhance the power-domain flexibility via NOMA.
In this paper, we investigate the synergy and mutual influence of beam hopping and NOMA.
We jointly optimize power allocation, beam scheduling, and terminal-timeslot assignment to minimize the gap between requested traffic demand and offered capacity.
In the solution development, we formally prove the NP-hardness of the optimization problem.
Next, we develop a bounding scheme to tightly gauge the global optimum and propose a suboptimal algorithm to enable efficient resource assignment.
Numerical results demonstrate the benefits of combining NOMA and BH, and validate the superiority of the proposed BH-NOMA schemes over benchmarks.
\end{abstract}



\section{Introduction}

\IEEEPARstart{I}{n} conventional multi-beam satellite systems, all beams are simultaneously illuminated and on-board resources are pre-assigned before launch due to limited flexibility and capability in satellite payloads \cite{maral2020satellite}.
While this design is efficient for static and uniform traffic patterns, the evolution of data services leads to highly dynamic and spatially non-uniform traffic.
In this case, the efficiency of resource utilization is low and the system fails to adapt to heterogeneous traffic distribution over the coverage area \cite{kodheli2020satellite}.
With the development of advanced satellite payloads, more attention has been drawn to flexible on-board resource allocation (e.g., power, bandwidth) to embrace the dramatic growth of data traffic and the uneven traffic distribution \cite{kodheli2020satellite,cocco2017radio}.

Beam hopping (BH) is a promising technique to enhance the flexibility of resource management by selectively and sequentially activating or deactivating beams \cite{kodheli2020satellite,angeletti2006beam,freedman2015beam}.
The benefits of BH are from the following aspects.
First, in a BH system, beam scheduling (or beam illumination pattern design) is optimized based on the requested traffic such that unmet and unused capacity can be reduced \cite{angeletti2006beam,freedman2015beam}.
Second, without illuminating all the beams together, the required number of radio-frequency chains is smaller, thus power consumption and payload mass are reduced \cite{freedman2015beam}.
Third, spatially induced co-channel interference can be alleviated by illuminating the beams that are distant from each other \cite{kibria2019precoded,lei2011multibeam}.
In the DVB-S2X standard \cite{dvbs2x}, a super-frame format to facilitate BH implementation and performance enhancement has been specified.
In the literature, BH has been applied in different scenarios, e.g., load balancing networks \cite{cao2019qos}, cognitive satellite networks \cite{zuo2018resource}, and ultra-dense LEO systems \cite{9473538}.  

\subsection{Related Works}
To improve the performance of BH, a majority of works focus on how to design efficient approaches to decide beam-timeslot scheduling. 
In \cite{angeletti2006beam}, a genetic algorithm was adopted to determine beam illumination patterns.
In \cite{zuo2018resource}, a resource allocation problem for cognitive BH systems was studied.
The authors decomposed the problem and proposed low-complexity approaches.
In \cite{alegre2012offered}, the authors designed two iterative BH approaches based on minimum co-channel interference and maximum signal-to-interference-plus-noise ratio (SINR). 
The authors in \cite{chen2021satellite} studied resource allocation for a novel satellite system where conventional BH is combined with cluster hopping. 
Considering the benefits of machine learning techniques, the authors in \cite{hu2020dynamic} and \cite{lei2020beam} proposed resource allocation schemes assisted by deep reinforcement learning and deep learning, respectively.

Compared to conventional orthogonal multiple access (OMA), non-orthogonal multiple access (NOMA) can achieve higher spectral efficiency and serve more terminals \cite{yan2019application}.
Beyond terrestrial systems, it is natural to investigate how NOMA can help to improve the performance for multi-beam satellite systems, e.g.,  \cite{perez2019non,wang2020noma,chu2021robust,jiao2020network}.
The authors in \cite{perez2019non} studied the cooperation between NOMA and precoding in a multi-beam satellite system. 
In \cite{wang2020noma}, joint optimization of power allocation, decoding orders, and terminal-timeslot assignment in NOMA-enabled multi-beam satellite systems was studied.
To mitigate channel-phase uncertainty effects, two robust beamforming schemes were provided in \cite{chu2021robust} to minimize power consumption for delivering satellite internet-of-things services.
In \cite{jiao2020network}, the authors jointly optimized power allocation and network stability to maximize long-term average capacity for NOMA-based satellite internet-of-things systems. 
NOMA has shown superiorities in enhancing spectral efficiency and improving the performance of practical metrics for multi-beam satellite systems in the literature.

Considering the individual benefits from BH and NOMA, we are motivated to investigate how to exploit the joint advantages of these two techniques and optimize resource allocation for BH-NOMA systems. 
In the literature, the joint scheme of BH and NOMA is studied to a limited extent.
The potential synergies of NOMA and BH were firstly studied in our previous work \cite{wang2021joint}, where we considered a simplified problem without joint optimization, and focused on performance evaluation in order to verify the initial synergy between BH and NOMA.

\subsection{Motivations and Contributions}
In general, joint resource optimization for BH-NOMA systems typically leads to a combinatorial optimization problem.
In some cases, the optimum might not be achievable for large-scale instances due to unaffordable complexity and time, e.g., branch-and-bound approach in solving large-scale integer linear programming problems \cite{lei2020beam,Kisseleff2020radio}.
For some difficult problems, the optimum might even be unknown for small or medium cases, e.g., unknown optimum in solving mixed-integer non-convex programming (MINCP) problems \cite{wang2021joint}.
It is therefore of importance to: 1) Identify how difficult the resource optimization problem is; 2) Provide a tight bound for the optimum; 3) Properly benchmark the developed suboptimal solutions. 

In this paper, we investigate joint optimization for the considered BH-NOMA scheme to enhance the performance gain by optimizing power allocation, beam scheduling, and terminal-timeslot assignment.
We apply BH to selectively and sequentially activate beams over timeslots.
NOMA is then implemented within each active beam to further improve the spectral efficiency.
Beyond state-of-the-art and compared to \cite{wang2021joint}, the main contributions are summarized as follows:
\begin{itemize}
\item We formulate a resource allocation problem to minimize the gap between offered capacity and requested traffic, leveraging by BH and NOMA.
The work, together with  \cite{wang2021joint}, provides an early-attempt investigation for BH-NOMA systems.
\item We formally prove the NP-hardness of the joint BH-NOMA optimization problem and outline the mutual influence between BH and NOMA. 
We investigate the problem's insights by developing theoretical analysis. 
\item To gauge the unknown global optimum, we design an effective bounding scheme.
In the upper-bound approach (UBA), we develop an iterative near-optimal algorithm.
In the lower-bound approach (LBA), we first resolve the problem's non-convexity by simplifying the estimation of inter-beam interference.
Then we construct a mixed-integer conic programming (MICP) problem to approximate the original problem.
\item We design an efficient suboptimal algorithm for joint power allocation, beam scheduling, and terminal-timeslot assignment (E-JPBT) to overcome the high complexity in UBA and provide feasible solutions for large-scale instances. 
\item The numerical results validate the benefits of jointly considering BH and NOMA, and the tightness of the bounds in gauging optimality.
We demonstrate the superiority of the proposed BH-NOMA schemes in matching offered capacity to requested traffic compared to benchmarks.
\end{itemize}

The remainders of the paper are organized as follows:
A multi-beam system model with the coexistence of BH and NOMA is illustrated in Section II.
In Section III, we formulate a joint resource optimization problem and provide theoretical analysis of the problem.
The procedures of UBA and LBA are elaborated in Section IV and Section V, respectively.
In Section VI, we present the details of designing E-JPBT.
The performance of the proposed algorithms is evaluated and discussed in Section VII.
Section VIII concludes the paper. 

Some notations are defined as follows:
The operator $|\cdot|$ denotes the absolute value of a complex number or the cardinality of a set.
$f(x;y)$ represents the function $f(x,y)$ with given $y$.
$[x]^+$ is equivalent to the calculation of $\max\lbrace x,0\rbrace$.
The operation of $\mathcal{X}\times\mathcal{Y}=\{(x,y)|x\in\mathcal{X},y\in\mathcal{Y}\}$ denotes the Cartesian product of two sets.

\section{System Model}

\newcommand{\tabincell}[2]{\begin{tabular}{@{}#1@{}}#2\end{tabular}}  

We consider a geostationary earth orbit (GEO) satellite system which provides services to fixed ground terminals via forward links. 
The satellite generates $B$ spot beams to cover the targeted area.
We denote $\mathcal{B}$ as the set of the beams.
Let $\mathcal{K}$ and $\mathcal{K}_b$ represent the set of terminals in the system and in the $b$-th beam, respectively.
Note that users are assigned to beams implicitly based on their geographical coordinates \cite{esa}, e.g., user $k$ located within the 4.3 dB contour of the $b$-th beam's coverage area belonging to $\mathcal{K}_b$.
Each terminal is equipped with a single directional antenna.
All the beams share the same frequency band, i.e., 1-color frequency-reuse pattern.

\begin{figure}[!htp]
\centering
\includegraphics[scale=0.5]{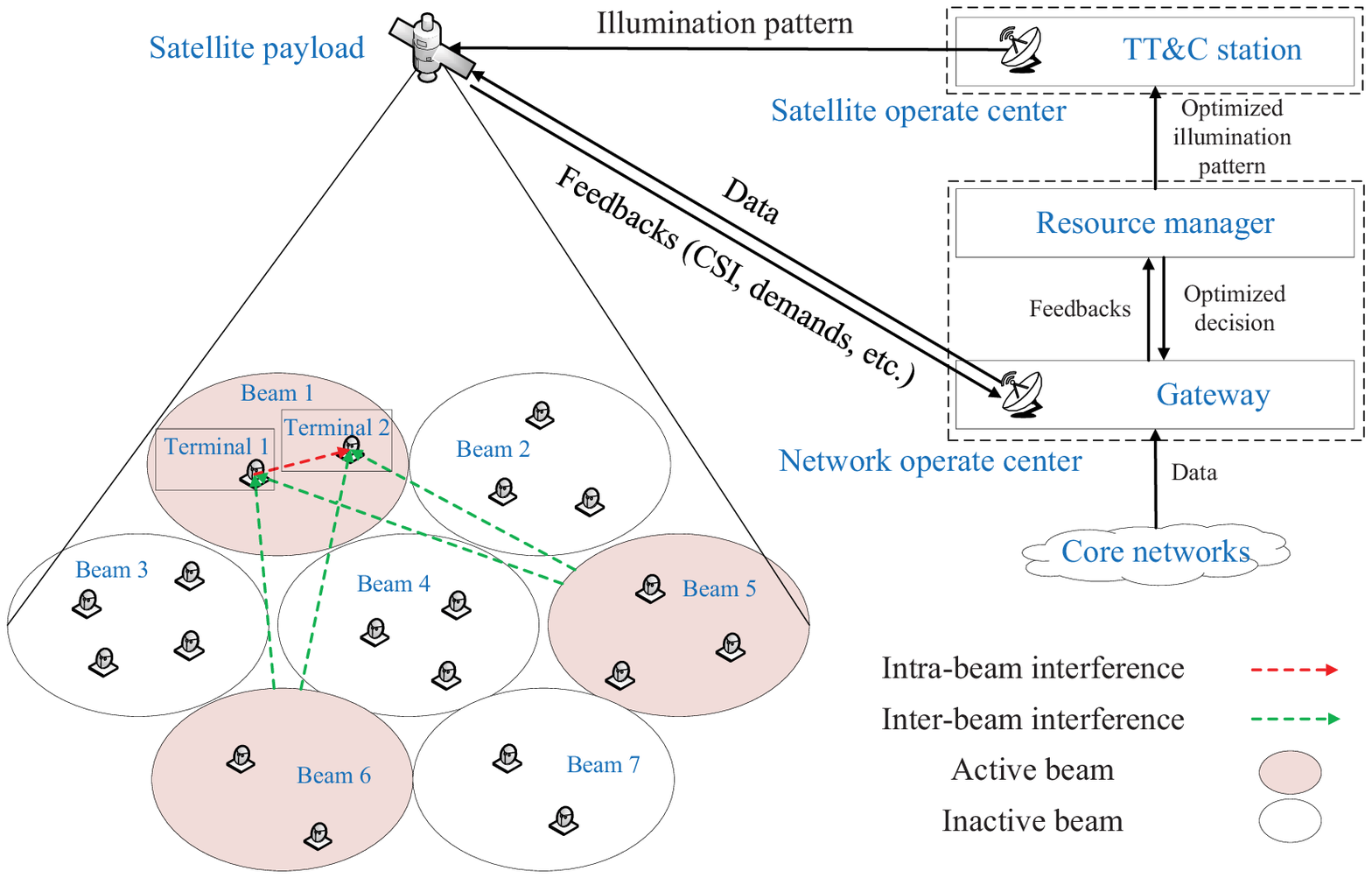}
\captionsetup{font={footnotesize}} 
\caption{An illustrative scenario of the considered BH-NOMA system. Three beams are activated simultaneously according to the BH design. By applying NOMA in beam 1, terminal 1 at the beam center with better channel gain only receives inter-beam interference from other two active beams, while terminal 2 at the beam edge with worse channel gain receives both intra-beam and inter-beam interference.}
\label{fig:sm2}
\end{figure}

The architecture of the considered multi-beam satellite system is depicted in Fig. \ref{fig:sm2}.
The telemetry, tracking, and command (TT\&C) station is part of the satellite operation center whereas the gateway and the resource manager are part of the network operation center \cite{kodheli2020satellite}.
The gateway collects information from ground terminals, e.g., traffic demands and channel status, via return links.
Based on the feedbacks, the resource manager optimizes the beam illumination pattern and NOMA-based resource allocation.
The decisions are communicated to the satellite payload via the TT\&C station and to the gateway \cite{lei2020beam,kodheli2020satellite}.
The TT\&C station is responsible for the synchronization among beams during the BH process \cite{panthi2017beam}.
Following the resource-allocation decisions, the gateway delivers data from the core networks to the satellite payload.
The bent-pipe transparent payload performs as a transceiver to relay data from the gateway to ground terminals.
The satellite payload is assumed to be equipped with switching matrix and digital transparent processors to enable BH and power distribution among different active beams, respectively \cite{kodheli2020satellite,maral2020satellite}.  

In the system, BH illuminates no more than $B_0$ ($B_0<B$) beams at each timeslot due to payload architecture limitations.
A scheduling period consists of $T$ timeslots, defined as a BH window.
Denote $\mathcal{T}$ as the set of the timeslots.
In an active beam, NOMA is adopted to multiplex one or more terminals in a timeslot.
The signals intended for the scheduled terminals in one beam are superimposed with different power per targeted terminal.
We denote $p_{kt}$ as the transmit power for terminal $k$ at timeslot $t$.
The SINR of terminal $k$ in beam $b$ at timeslot $t$ is derived as,
\begin{equation}
\gamma_{kt}=\frac{|h_{bk}|^2p_{kt}}{\underbrace{\sum\limits_{k{'}\in\mathcal{K}_b\setminus\{k\}\atop k'<k}|h_{bk}|^2p_{k't}}_{\textrm{intra-beam interference}}+\underbrace{\sum\limits_{b'\in\mathcal{B}\setminus\{b\}}\sum\limits_{k{'}\in\mathcal{K}_{b'}}|h_{b'k}|^2p_{k't}}_{\textrm{inter-beam interference}}+\sigma^2},
\label{sinr}
\end{equation}
where $\sigma^2$ represents the noise power.
We denote $|h_{bk}|^2$ as the channel gain from the $b$-th satellite antenna to the $k$-th terminal (assuming consistent indexes between antennas and beams).
We use a widely-adopted channel model in multi-beam satellite systems \cite{lei2020beam,zuo2018resource,kibria2019precoded,wang2020noma,chu2021robust}, which is derived as,
\begin{equation}
|h_{bk}|^2=\frac{G^{\mathrm{tx}}_{bk}G_{k}^{\mathrm{rx}}}{\kappa T^{\mathrm{noise}} W} \left(\frac{c}{4\pi d_{k} f^{\mathrm{fr}}}\right)^2,
\end{equation}
where $G_{bk}^{\mathrm{tx}}$ is the transmit antenna gain from the $b$-th antenna to terminal $k$. 
$G_{k}^{\mathrm{rx}}$ denotes the receive antenna gain of terminal $k$.
The term $\kappa T^{\mathrm{noise}} W$ represents the distribution of noise, where $\kappa$, $T^{\mathrm{noise}}$, and $W$ denote the Boltzmann constant, the noise temperature of the receiver, and the carrier bandwidth, respectively.
The term $\left(\frac{c}{4\pi d_{k} f^{\mathrm{fr}}}\right)^2$ is the free-space propagation loss, where $d_{k}$, $f^{\mathrm{fr}}$, and $c$ denote the distance between the satellite and terminal $k$, the frequency, and the light speed, respectively.
We consider that the channel gains are static within $T$ timeslots and updated every $T$ timeslots.

The intra-beam interference and inter-beam interference are denoted as the first term and the second term of the denominator in \eqref{sinr}, respectively.
We assume consistent indexes between terminals and the descending order of channel gains, e.g., two terminals $k$ and $k'$ in $\mathcal{K}_b$, where $k'<k$ and $|h_{bk'}|^2>|h_{bk}|^2$.
In this case, $k'$ performs successive interference cancellation (SIC) to decode and remove $k$'s signals whereas $k$ treats $k'$'s signals as noise.
We remark that, to facilitate the analysis, we assume that the channel coefficients satisfy the conditions derived in \cite{you2018resource} (Lemma 1), such that determining the decoding order in each beam is independent of the beams' transmit power and inter-beam interference.
The available rate of terminal $k$ at timeslot $t$ is,
\begin{equation}
R_{kt}=W\log_2(1+\gamma_{kt}),
\label{r}
\end{equation}
where $W$ is the bandwidth for the carrier (single carrier per beam).
The total offered capacity of terminal $k$ is,
\begin{equation} 
R_{k}=\sum_{t\in\mathcal{T}}R_{kt}.\label{capacity}
\end{equation}

\section{Problem Formulation and Analysis}

\subsection{Problem Formulation}
We formulate an optimization problem to minimize the gap between offered capacity and requested traffic by jointly optimizing power allocation, beam scheduling, and terminal-timeslot assignment.
The variables are defined as:
\begin{align}
&p_{kt}\geq 0,\textrm{ transmit power for terminal $k$ at timeslot $t$;}\notag\\
&\alpha_{bt}=
\begin{cases}
1,\textrm{ beam $b$ is illuminated at timeslot $t$,}\\
0,\textrm{ otherwise;}
\end{cases}\notag\\
&\beta_{kt}=
\begin{cases}
1,\textrm{ terminal $k$ is assigned to timeslot $t$,}\\
0,\textrm{ otherwise.}
\end{cases}\notag
\end{align}
Denote $D_{k}$ as the requested traffic demand (in bps) of terminal $k$ over a scheduling period.
We apply a widely-adopted metric, $(R_k-D_k)^2$, to measure the capacity-demand mismatch of terminal $k$  \cite{lei2011multibeam,cocco2017radio}.
The objective function captures the average mismatch level among $K$ terminals.
The problem is formulated as:
\begin{subequations}
\begin{align}
    \mathcal{P}_0: &\min_{\alpha_{bt},\beta_{kt},p_{kt}} \sum_{k\in\mathcal{K}}(R_{k}-D_{k})^2\label{po1}\\ 
    \,\,\,\,\mbox{s.t.}\,\,\,\,
    &\sum_{k\in\mathcal{K}_b}p_{kt}\leq P,\forall b\in\mathcal{B}, \forall t\in\mathcal{T},\label{pc1}\\
    &\sum_{b\in\mathcal{B}}\alpha_{bt}\leq B_0,\forall t\in\mathcal{T},\label{pc2}\\
    &\sum_{k\in\mathcal{K}_b}\beta_{kt}\leq K_0\alpha_{bt},\forall b\in\mathcal{B},\forall t\in\mathcal{T},\label{pc3}\\
    &p_{kt}\leq P\beta_{kt},\forall k\in\mathcal{K}, \forall t\in\mathcal{T},\label{pc4}\\
    &R_{k}\geq R^{\min}_{k},\forall k\in\mathcal{K},\label{pc5}\\
    &\alpha_{bt}+\alpha_{b^{'}t}\leq 1, b\neq b^{'}, \forall (b,b^{'})\in{\Omega},\forall t\in\mathcal{T}.\label{pc6}
\end{align}
\end{subequations}
In \eqref{pc1}, the total transmit power of terminals in each beam at each timeslot is no larger than the beam power budget $P$.
In \eqref{pc2}, no more than $B_0$ beams can be illuminated at each timeslot. 
Constraints \eqref{pc3} confine that no more than $K_0$ terminals can be allocated to a timeslot.
No terminal will be scheduled in an inactive beam.
Constraints \eqref{pc4} connect $p_{kt}$ and $\beta_{kt}$, where $p_{kt}$ is confined to zero if $\beta_{kt}=0$, otherwise, $p_{kt}\leq P$.
In \eqref{pc5}, the rate of each terminal should meet the minimum-rate requirement to maintain a certain level of fairness among terminals.
Usually, the minimum rate is smaller than the requested traffic demand, i.e., $R_{k}^{\min}<D_{k}$.
In \eqref{pc6}, we introduce $\Omega\subset\mathcal{B}\times\mathcal{B}$ as a set to include all the undesired beam pairs, e.g., adjacent beams with strong interference.
If a beam pair $\{b,b'\}\in\Omega$, beam $b$ or $b'$ can be illuminated alone or grouped with other beams, e.g., illuminating beam $b$ and $b''$ in timeslot $t$, but beam $b$ and $b'$ cannot be activated together in the same timeslot because $\alpha_{bt}+\alpha_{b't}=2$ violates \eqref{pc6}.

In $\mathcal{P}_0$, the performance and optimization decisions in BH and NOMA are coupled with each other.
In general, jointly optimizing the two components is challenging.
Determining NOMA resource allocation in each beam depends on the outcome of BH design, whereas achieving a high-quality BH scheme requires appropriate decisions from NOMA resource allocation.
On the one hand, BH design is of importance to  resource allocation in NOMA.
When a set of inappropriate beams with strong interference are activated, this can possibly result in degraded performance, e.g., low data rates per timeslot.
As a consequence, each terminal might need to be assigned with more power to satisfy its demand or scheduled to more timeslots (thus suggests more timeslots consumed in total for all the terminals), which typically leads to a more complicated problem with a larger dimension and more sensitive to the feasibility issue in NOMA.

On the other hand, the decisions made in NOMA can in its turn influence the BH design. 
When an optimal power and terminal-timeslot allocation can be obtained in NOMA, as a result, each active beam radiates less inter-beam interference to each other compared to a suboptimal NOMA solution, and some beams can be activated with fewer timeslots due to the higher rate achieved per timeslot, which can greatly ease the BH design.
Towards an overall high-quality solution for BH-NOMA systems, it is necessary to capture this mutual dependence and iteratively improve the overall performance in algorithmic design.

\subsection{Complexity Analysis in Solving $\mathcal{P}_0$}
$\mathcal{P}_0$ is an MINCP problem \cite{boyd2004convex} due to the nonlinear and nonconvex functions in \eqref{po1} and \eqref{pc5}, and the presence of binary variables $\alpha_{bt}$ and $\beta_{kt}$.
Solving an MINCP can be challenging in general.
We further identify the intractability of $\mathcal{P}_0$ by proving the NP-completeness for its decision-version problem (or feasibility-check problem) and the NP-hardness for the optimization problem in Lemma 1 and Theorem 1, respectively.
The decision-version problem of $\mathcal{P}_0$ is defined as a true-or-false problem to check if there exists a feasible solution \cite{liu2013complexity}.
If the decision version of $\mathcal{P}_0$ is NP-complete, then the optimization problem $\mathcal{P}_0$ is NP-hard \cite{hartmanis1982computers}, because solving $\mathcal{P}_0$ is no easier than solving its decision version.
The former needs to obtain optimal solutions, whereas the latter only needs to offer a yes-or-no answer for feasibility check.

\textit{Lemma 1: The decision-version (feasibility-check) problem of $\mathcal{P}_0$ is NP-complete.}

\begin{proof}
Please refer to Appendix \ref{lemma1}.
\end{proof}

Based on Lemma 1, the NP-hardness of $\mathcal{P}_0$ can be therefore concluded.

\textit{Theorem 1: $\mathcal{P}_0$ is NP-hard.}

Being aware of the NP-hardness of $\mathcal{P}_0$ and the coupling effects between BH and NOMA, it is challenging to solve the original problem directly.
Instead, we fix the binary variables and provide theoretical analysis of how to deal with the remaining problem.

It is worth noting that, even with the fixed binary variables $\alpha_{bt}$ and $\beta_{kt}$, the remaining power allocation problem, shown as in $\mathcal{P}_1$, is still non-convex \cite{boyd2004convex}.
\begin{subequations}
\begin{align}
    \mathcal{P}_1: &\min_{p_{kt}} \sum_{k\in\mathcal{K}}(R_{k}-D_{k})^2\\ 
    \,\,\,\,\mbox{s.t.}\,\,\,\,
    &\eqref{pc1}, \eqref{pc5},
\end{align}
\end{subequations}
where $p_{kt}\geq 0$ for $\beta_{kt}=1$ and $p_{kt}=0$ for $\beta_{kt}=0$. 
We introduce auxiliary variables $\delta_{k}$ and equivalently convert $\mathcal{P}_1$ as,
\begin{subequations}
\begin{align}
    \mathcal{P}'_1: &\min_{p_{kt},\delta_k} \sum_{k\in\mathcal{K}}\delta_k^2\\ 
    \,\,\,\,\mbox{s.t.}\,\,\,\,
    &\eqref{pc1}, \eqref{pc5},\\
    &-\delta_k\leq R_k-D_k\leq \delta_k,\forall k\in\mathcal{K}.\label{pco}
\end{align}
\end{subequations}
At the optimum, $R_k-D_k$ is equal to either $-\delta_k$ or $\delta_k$, $\forall k\in\mathcal{K}$, where $\delta_k\geq 0$.
In the following proposition, we prove that $R_k\leq D_k$ at the optimum, which can simplify $\mathcal{P}'_1$.

\textit{Proposition 1:}
\textit{At the optimum of $\mathcal{P}_1$, $R_{k}\leq D_{k}$, $\forall k\in\mathcal{K}$.}

\begin{proof}
Please refer to Appendix \ref{theorem2}.
\end{proof}

With Proposition 1, constraints \eqref{pco} can be converted into,
\begin{equation}
-\delta_k\leq R_k-D_k, \forall k\in\mathcal{K},\label{rddelta}
\end{equation}
which indicates that $R_k-D_k=-\delta_k$ at the optimum.
In spite of the problem conversion, we observe that solving $\mathcal{P}'_1$ remains challenging due to the nonconvexity of the logarithmic-fractional composite expressions in the $R$-functions \cite{boyd2004convex}.
A widely-adopted approach to address the fractional nonconvex function is to decouple the numerator and denominator, and transform it into a series of convex problems, e.g., Dinkelbach's transform \cite{zappone2017globally}, and quadratic transform \cite{shen2018fractional}.
Compared to conventional Dinkelbach's transform, quadratic transform has shown advantages in tackling multi-ratio fractional programming by building the equivalence of the objectives between the primal and the transformed problem \cite{shen2018fractional}.
Besides, quadratic transform has proven its competitiveness compared to conventional successive convex approximation method \cite{wang2012successive} in power control \cite{shen2018fractional}. 
By applying quadratic transform \cite{shen2018fractional}, we convert $R_{kt}$ from fractional format to the following,
\begin{align}
&f^R_{kt}(\theta_{kt},p_{kt})=\log\left(1+2\theta_{kt}\sqrt{|h_{bk}|^2p_{kt}}\right.\notag\\&\left.-\theta_{kt}^2\!\!\left(\!\!\sum\limits_{k{'}\in\mathcal{K}_b\setminus\{k\}\atop k'<k}\!\!\!\!\!\!\!|h_{bk}|^2p_{k't}+\!\!\!\sum\limits_{b'\in\mathcal{B}\setminus\{b\}}\sum\limits_{k{'}\in\mathcal{K}_{b'}}\!\!\!|h_{b'k}|^2p_{k't}+\sigma^2\right)\!\!\right)\!\!,
\label{ldq}
\end{align}
where $\theta_{kt}\geq 0$ is the auxiliary variable.
With fixed $\theta_{kt}$, $f^R_{kt}$ is a concave function according to the basis of convex preservation for composite functions \cite{boyd2004convex}.
Then $\mathcal{P}'_1$ is rewritten as the following,
\begin{subequations}
\begin{align}
    \mathcal{P}_2: &\min_{\theta_{kt},p_{kt},\delta_k} \sum_{k\in\mathcal{K}}\delta_k^2\\ 
    \,\,\,\,\mbox{s.t.}\,\,\,\,
    &\eqref{pc1}, \\
    &\sum_{t\in\mathcal{T}}f^R_{kt}(\theta_{kt},p_{kt})\geq R^{\min}_{k},\forall k\in\mathcal{K},\label{pc53}\\
    & \sum_{t\in\mathcal{T}}f^R_{kt}(\theta_{kt},p_{kt})-D_k\geq -\delta_k,\forall k\in\mathcal{K}.\label{pc63}
\end{align}
\end{subequations}
$\mathcal{P}_2$ is nonconvex in general, but can become convex when $\theta_{kt}$ is fixed, 
which enables an iterative approach to optimize $p_{kt}$ with fixed $\theta_{kt}$ by solving the convex problem and updating $\theta_{kt}$ under fixed $p_{kt}$.

\section{An Iterative Approach for Upper Bound}
In this section, we propose UBA algorithm to obtain an upper bound (a feasible suboptimal solution) for $\mathcal{P}_0$.
In UBA, we optimize power allocation with fixed integer variables and iteratively update beam scheduling and terminal-timeslot assignment to progressively improve the performance.

\subsection{Power Allocation with Fixed Integer Solution}
The considered iterative algorithm for solving $\mathcal{P}_2$ is summarized in Alg. \ref{alg:fp}.
In each iteration, line 2 and line 3 describe the procedures of alternatively update $\theta_{kt}$ and $p_{kt}$, respectively, where $\theta_{kt}$ is updated with fixed power allocation and $p_{kt}$ is optimized given $\theta_{kt}$. 
With fixed $p_{kt}$, the optimal $\theta_{kt}$ is derived by \cite{shen2018fractional}, 
\begin{equation}
\theta_{kt}=\frac{\sqrt{|h_{bk}|^2p_{kt}}}{\sum\limits_{k{'}\in\mathcal{K}_b\setminus\{k\}\atop k'<k}|h_{bk}|^2p_{k't}+\sum\limits_{b'\in\mathcal{B}\setminus\{b\}}\sum\limits_{k{'}\in\mathcal{K}_{b'}}|h_{b'k}|^2p_{k't}+\sigma^2}.
\label{theta_opt}
\end{equation}
With fixed $\theta_{kt}$, $\mathcal{P}_2$ becomes convex.
The optimum can be obtained by conventional algorithms, e.g., interior-point method \cite{schurr2009polynomial}.
Based on the theoretical results in \cite{shen2018fractional}, we conclude that the iterative process in lines 1-4 converges to a stationary point.
At the end of convergence, there may exist terminals with $\sum_{t\in\mathcal{T}}f^R_{kt}> D_k$.
According to the conclusion of Proposition 1, a post process in lines 6-8 is performed for these terminals by solving the following equations,
\begin{equation}
R_{k}=D_k, \forall k\in\mathcal{K}^*,\label{requd}
\end{equation}
where $\mathcal{K}^*$ includes the terminals with $R_k\geq D_k$.
The nonlinear equations can be solved via the Levenberg-Marquardt method \cite{zhao2016global}.

\begin{algorithm}[t]
\caption{Iterative approach for power allocation}
\label{alg:fp}
\begin{algorithmic}[1]
\REQUIRE 
feasible $p_{kt}$ and $\delta_{k}$.
\REPEAT
\STATE Update $\theta_{kt}$ by \eqref{theta_opt}. 
\STATE Optimize $p_{kt}$ and $\delta_{k}$ by solving $\mathcal{P}_2$.
\UNTIL{convergence}
\STATE Calculate $R_{k}$ by \eqref{sinr}, \eqref{r}, \eqref{capacity}.
\IF{there exists terminals with $R_k>D_k$}
\STATE Solve nonlinear equations in \eqref{requd}.
\ENDIF
\ENSURE 
optimized $p_{kt}$, $\delta_{k}$.
\end{algorithmic}
\end{algorithm}

%
%
%

The complexity of Alg. \ref{alg:fp} mainly falls into the optimization process in line 3 and solving nonlinear equations in line 7.
For optimizing $p_{kt}$ in line 3, we apply interior-point method to solve $\mathcal{P}_2$ with the complexity of $\mathcal{O}(\psi\log(\frac{1}{\varepsilon}))$ \cite{schurr2009polynomial}, where $\psi>0$ is the parameter for self-concordant barrier and $\varepsilon>0$ is the precision \cite{schurr2009polynomial}.
The complexity of solving nonlinear equations in line 7 is $\mathcal{O}({\varrho^{-2}})$.
Here, $\varrho>0$ satisfies $||\mathbf{J}^T\mathbf{F}||\leq\varrho$, where $\mathbf{F}=\mathbf{0}$ is the nonlinear equations and $\mathbf{J}$ is the corresponding Jacobian matrix  \cite{zhao2016global}.
The complexity of Alg. \ref{alg:fp} is therefore $\mathcal{O}(\max\{N\psi\log(\frac{1}{\varepsilon}),\varrho^{-2}\})$, where $N$ is the maximum number of iterations.

\subsection{Beam Scheduling and Terminal-Timeslot Assignment}
Next, we jointly optimize beam scheduling and terminal-timeslot assignment to improve the performance iteratively. 
Some approaches, e.g., exhaustive search method, or branch and bound \cite{lee2011mixed}, are capable of obtaining the optimal or near-optimal integer solution but at the expense of unaffordable computational complexity.
To avoid exponential-time complexity, we provide a scheme based on matching theory to decrease the capacity-demand gap iteratively.

The optimization of integer solutions can be viewed as two many-to-many matching problems \cite{di2016sub}, i.e., beam-to-timeslot matching and terminal-to-timeslot matching.
Define $\mathcal{M}$ and $\mathcal{N}$ as the sets containing tuples of beam-to-timeslot and terminal-to-timeslot matching, respectively.
Denote $y(\mathcal{M},\mathcal{N})$ as the objective value obtained by Alg. \ref{alg:fp} under $\mathcal{M}$ and $\mathcal{N}$, where $\mathcal{M}$ and $\mathcal{N}$ are constructed based on $\alpha_{bt}$ and $\beta_{kt}$, respectively.
For instance, if $\alpha_{bt}=1$, then $(b,t)\in\mathcal{M}$, otherwise, $(b,t)\notin\mathcal{M}$.
The many-to-many matching problem can be solved via swap \cite{di2016sub}.
Consider a set $\mathcal{M}_1$, where $(b,t)\in\mathcal{M}_1$ but $(b',t')\notin\mathcal{M}_1$.
We define a swap $\mathcal{S}^{bt}_{b't'}$ as the operation of converting $\mathcal{M}_1$ into $\mathcal{M}_2$ by removing $(b,t)$ and adding $(b',t')$, i.e.,  setting $\alpha_{bt}=0$ and $\alpha_{b't'}=1$.
A swap happens if $\mathcal{M}_1$ and $\mathcal{M}_2$ meet the following conditions:
\begin{align}
&\textrm{1. $\mathcal{M}_2$ satisfies constraints \eqref{pc2} and \eqref{pc6};}\label{con12}\\
&\textrm{2. $y(\mathcal{M}_1,\mathcal{N})> y(\mathcal{M}_2,\mathcal{N})$.}\label{con13}
\end{align}
Define $\mathfrak{S}$ as the set containing all the possible $\mathcal{S}_{b't'}^{bt}$.
\begin{algorithm}[t]
\caption{UBA}
\label{alg:sm}
\begin{algorithmic}[1]
\REQUIRE 
 Feasible $\tilde{p}_{kt}$, $\tilde{\alpha}_{bt}$, and $\tilde{\beta}_{kt}$ (corresponding to $\tilde{\mathcal{M}}$ and $\tilde{\mathcal{N}}$).
\REPEAT
\STATE Construct $\mathfrak{S}$ based on $\tilde{\alpha}_{bt}$ and \eqref{con12}.
\IF{$\mathfrak{S}\neq\varnothing$}
\STATE Select a swap $\mathcal{S}^{bt}_{b't'}$ and construct $\mathcal{M}$ with $(b',t')$.
\STATE Optimize $p_{kt}$ under $\mathcal{M}$ via Alg. \ref{alg:fp}. 
\ELSE
\STATE {UBA terminates.} 
\ENDIF 
\IF{$\tilde{\mathcal{M}}$ and $\mathcal{M}$ do not satisfy \eqref{con13}}
\STATE Remove $\mathcal{S}^{bt}_{b't'}$ from $\mathfrak{S}$.
\STATE Move to line 3.
\ENDIF
\STATE Let $\tilde{\mathcal{M}}=\mathcal{M}$ and $\tilde{p}_{kt}=p_{kt}$ and update $\tilde{\alpha}_{bt}$, $\tilde{\beta}_{kt}$, and $\tilde{\mathcal{N}}$. Construct $\bar{\mathfrak{S}}$ based on $\tilde{\beta}_{kt}$,  \eqref{con22}, and \eqref{con23}.
\IF{$\bar{\mathfrak{S}}\neq\varnothing$}
\STATE Select a swap $\bar{\mathcal{S}}^{kt}_{k't'}$ and construct $\mathcal{N}$ with $(k',t')$.
\STATE Optimize $p_{kt}$ under $\mathcal{N}$ via Alg. \ref{alg:fp}. 
\ENDIF 
\IF{$\tilde{\mathcal{N}}$ and $\mathcal{N}$ do not satisfy \eqref{con24}}
\STATE Remove $\bar{\mathcal{S}}^{kt}_{k't'}$ from $\bar{\mathfrak{S}}$.
\STATE Move to line 14.
\ENDIF
\STATE Let $\tilde{\mathcal{N}}=\mathcal{N}$ and $\tilde{p}_{kt}=p_{kt}$ and update $\tilde{\beta}_{kt}$.
\UNTIL{the number of iterations reaches $N'$ or $R_k=D_k$ $\forall k\in\mathcal{K}$}
\ENSURE 
$\tilde{p}_{kt}$, $\tilde{\alpha}_{bt}$, $\tilde{\beta}_{kt}$.
\end{algorithmic}
\end{algorithm}

Analogous to beam-to-timeslot swap, we consider a set $\mathcal{N}_1$, where $(k,t)\in\mathcal{N}_1$ but $(k',t')\notin\mathcal{N}_1$.
Define a swap $\bar{\mathcal{S}}^{kt}_{k't'}$ as the operation of converting $\mathcal{N}_1$ into $\mathcal{N}_2$ by removing $(k,t)$ and introducing $(k',t')$, i.e., setting $\beta_{kt}=0$ and $\beta_{k't'}=1$.
A swap occurs if $\mathcal{N}_1$ and $\mathcal{N}_2$ satisfy the following conditions:
\begin{align}
&\textrm{1. $\mathcal{N}_2$ satisfies constraints \eqref{pc3};}\label{con22}\\
&\textrm{2. $k, k^{'}\in\mathcal{K}_b$ and $\alpha_{bt}=\alpha_{bt^{'}}=1$;}\label{con23}\\
&\textrm{3. $y(\mathcal{M},\mathcal{N}_1)> y(\mathcal{M},\mathcal{N}_2)$.}\label{con24}
\end{align}
Define $\bar{\mathfrak{S}}$ as the set containing all the possible $\bar{\mathcal{S}}_{k't'}^{kt}$.

We summarize the procedure of UBA in Alg. \ref{alg:sm}.
Denote $N'$ as the maximum number of iterations.
Line 3 to line 13 represent the swap of beam-to-timeslot matching and line 14 to line 22 indicate the terminal-to-timeslot swap.
Remark that the algorithm starts to assign terminals to timeslots once beam-to-timeslot swap is executed.
The algorithm terminates when the number of iterations reaches $N'$, there is no more valid swap, or all the terminals are satisfied with demand.

In UBA, the complexity for each iteration consists of two parts, i.e., beam-to-timeslot swap and terminal-to-timeslot swap.
The numbers of all the possible beam-to-timeslot and terminal-to-timeslot swaps are at most $TB_0\times T(B-B_0)$  and $B_0\times TK_0\times T(K'-K_0)$, respectively, where $K'$ denotes the maximum number of terminals in each beam.
For each iteration, the worst case is to optimize power for all the swaps in $\mathfrak{S}$ and $\bar{\mathfrak{S}}$ by Alg. \ref{alg:fp}.
Thus the complexity of UBA is $\mathcal{O}(N'T^2B_0(B-B_0+K_0(K'-K_0))\max\{N\psi\log(\frac{1}{\varepsilon}),\varrho^{-2}\})$.
Remark that, in practice, the complexity can be largely reduced by eliminating a certain number of swaps that do not satisfy conditions \eqref{con12}, \eqref{con22}, and \eqref{con23}.

\section{An MICP Approximation Approach for Lower Bound}
In this section, we resolve $\mathcal{P}_0$'s non-convexity by intentionally simplifying the inter-beam interference, and construct an MICP formulation to enable a lower bound for $\mathcal{P}_0$.
We observe that, in some BH cases, the inter-beam interference may become negligible, e.g., illuminating two distant beams.
If this interference can be ignored, $\mathcal{P}_0$ becomes an MICP problem.
In this case, $p_{kt}$ for terminals in $\mathcal{K}_b$ can be derived as follows,
\begin{align}
&p_{1t}=\left(2^{\frac{R_{1t}}{W}}-1\right)\frac{\sigma^2}{|h_{b1}|^2},\notag\\
&p_{2t}=\left(2^{\frac{R_{2t}}{W}}-1\right)\left(p_{1t}+\frac{\sigma^2}{|h_{b2}|^2}\right),\notag\\
&\dots\notag\\
&p_{{K_b}t}=\left(2^{\frac{R_{{K_b}t}}{W}}-1\right)\left(\sum_{k'=1}^{{K_b}-1}p_{k't}+\frac{\sigma^2}{|h_{b{K_b}}|^2}\right).
\end{align}
Thus \eqref{pc1} can be rewritten in an equivalent expression as,
\begin{align}
&\sum_{k=1}^{{K}_b}\left(\frac{\sigma^2}{|h_{bk}|^2}-\frac{\sigma^2}{|h_{b(k-1)}|^2}\right)2^{\frac{\sum\limits_{k'=k}^{K_b}R_{k't}}{W}}-\frac{\sigma^2}{|h_{bK_b}|^2}\leq P,\notag\\
&\,\,\,\,\,\,\,\,\,\,\,\,\,\,\,\,\,\,\,\,\,\,\,\,\,\,\,\,\,\,\,\,\,\,\,\,\,\,\,\,\,\,\,\,\,\,\,\,\,\,\,\,\,\,\,\,\,\,\,\,\,\,\,\,\,\,\,\,\,\,\,\,\,\,\,\,\,\,\,\,\,\,\,\,\,\,\,\,\,\,\,\,\,\forall b\in\mathcal{B},\forall t\in\mathcal{T},\label{expr}
\end{align}
where we define that $\frac{\sigma^2}{|h_{b0}|^2}=0$.
Then $\mathcal{P}_0$ becomes,
\begin{subequations}
\begin{align}
\mathcal{P}_3: &\min_{\alpha_{bt},\beta_{kt},R_{kt},\delta_k}\sum_{k\in\mathcal{K}}\delta_k^2\label{obj3}\\
\,\,\,\,\mbox{s.t.}\,\,\,\,
&\eqref{expr}, \eqref{pc2}, \eqref{pc3},\eqref{pc5},\eqref{pc6},\eqref{pco},\\
&R_{kt}\leq R^{\max}\beta_{kt}, \forall k\in\mathcal{K}, \forall t\in\mathcal{T},\label{pcr}
\end{align}
\end{subequations}
where $R^{\max}$ is a constant, no smaller than all possible $R_{kt}$.
Constraints \eqref{pcr} connect $R$-variables with $\beta$-variables, where $R_{kt}=0$ if $\beta_{kt}=0$, otherwise, $R_{kt}\leq R^{\max}$.
$\mathcal{P}_3$ is an MICP with the presence of exponential cones in \eqref{expr} and the quadratic cones in \eqref{obj3}.
The optimum of $\mathcal{P}_3$ can be achieved by branch-and-bound or outer approximation approach \cite{lee2011mixed}.

Note that, since inter-beam interference has been intentionally removed, the optimum of $\mathcal{P}_3$ is no larger than  that of $\mathcal{P}_0$ and thus can be viewed as a lower bound of $\mathcal{P}_0$.

\textit{Remark 1:} 
The gap between the lower bound (the optimum of $\mathcal{P}_3$) and the optimum of $\mathcal{P}_0$ follows three cases:
\begin{itemize}
\item Zero gap: The optimum at $\mathcal{P}_0$ and $\mathcal{P}_3$ is equivalent if only one beam is illuminated at each timeslot since there is no inter-beam interference in the system.
\item Close-to-zero gap: $\mathcal{P}_3$ provides a close lower bound to the optimum of $\mathcal{P}_0$ if the level of inter-beam interference keeps low.
\item Large gap: In some cases, the lower bound becomes loose when inter-beam interference is strong. 
However, due to the inherent characteristics in BH optimization, only the beams with less mutual interference are preferred to be activated at the same timeslot.
Thus, this undesired issue can be avoided in a majority of cases.
\hfill $\square$
\end{itemize}

\section{An Efficient Algorithm for Joint Optimization}
UBA aims at providing a tight upper bound (or near-optimal solution) to $\mathcal{P}_0$ at the expense of high complexity.
To further reduce the computational complexity, we design a low-complexity approach, i.e., E-JPBT, to provide an efficient solution for $\mathcal{P}_0$.
The basic idea of E-JPBT is to divide the whole decision process of $\mathcal{P}_0$ into $T$ stages (or timeslots) and then solve a subproblem at each stage or timeslot.
To avoid directly tackling integer variables with large complexity, the subproblem for each timeslot is relaxed to a continuous problem, which can be solved by Alg. \ref{alg:fp}.

The residual demand for terminal $k$ at timeslot $t$ is, 
\begin{equation}
\bar{D}_k=D_{k}-\sum_{\tau=0}^{t-1}R_{k\tau},\label{resid}
\end{equation}
where $R_{0t}=0$.
At the $t$-th timeslot, the resource allocation problem for the current timeslot is expressed as,
\begin{subequations}
\begin{align}
\mathcal{P}_0(t): &\min_{\alpha_{bt},\beta_{kt},p_{kt}}\sum_{k\in\mathcal{K}}\left(R_{kt}-\bar{D}_{k}\right)^2\notag\\
&+\sum_{k\in\mathcal{K}}\phi_{k}\left[R^{\min}_{k}-R_{kt}-\sum_{\tau=0}^{t-1}R_{k\tau}\right]^+\\
\,\,\mbox{s.t.}\,\,
&\sum_{k\in\mathcal{K}_b}p_{kt}\leq 1,\forall b\in\mathcal{B},\label{pc1t}\\
&\sum_{b\in\mathcal{B}}\alpha_{bt}\leq B_0,\label{pc2t}\\
&\sum_{k\in\mathcal{K}_b}\beta_{kt}\leq K_0\alpha_{bt},\forall b\in\mathcal{B},\label{pc3t}\\
&p_{kt}\leq \beta_{kt},\forall k\in\mathcal{K}_b,\label{pc4t}\\&\alpha_{bt}+\alpha_{b^{'}t}\leq 1, b\neq b^{'}, \forall (b,b^{'})\in{\Omega},\label{pc6t}
\end{align}
\end{subequations}
where the second term of the objective is the penalty for constraints \eqref{pc5} and $\phi_{k}>0$ is the penalty factor.
The objective is penalized if the rate of terminal $k$ is lower than $R_{k}^{\min}$, which means more resources should be allocated to this terminal in the later timeslots.

Since $\mathcal{P}_0(t)$ is MINCP and is still challenging, we relax $\alpha_{bt}$ and $\beta_{kt}$ into continuous variables, i.e., $0\leq\bar{\alpha}_{bt}\leq 1$ and $0\leq\bar{\beta}_{kt}\leq 1$.
Then we convert the $R$-function into $f_{kt}^{R}(\theta_{kt},p_{kt})$ as expressed in \eqref{ldq} via quadratic transform.
$\mathcal{P}_0(t)$ can be reformulated as,
\begin{subequations}
\begin{align}
\mathcal{P}_4(t): &\min_{\bar{\alpha}_{bt},\bar{\beta}_{kt},p_{kt}, \theta_{kt}}\sum_{k\in\mathcal{K}}\left(f^R_{kt}-\bar{D}_{k}\right)^2\notag\\
&+\sum_{k\in\mathcal{K}}\phi_{k}\left[R^{\min}_{k}-f^R_{kt}-\sum_{\tau=0}^{t-1}f^R_{k\tau}\right]^+\\
\,\,\mbox{s.t.}\,\,
&\eqref{pc1t}, \eqref{pc2t}, \eqref{pc3t},\eqref{pc4t},\eqref{pc6t},
\end{align}
\end{subequations}
which can be solved via Alg. \ref{alg:fp}.
We define $\mathcal{B}_j\subseteq\mathcal{B}$ as the $j$-th beam group.
Note that the beam groups are constructed based on $\Omega$.
We schedule the beam group on the basis of $j^*=\arg\max_{j}\{\sum_{b\in\mathcal{B}_j}\bar{\alpha}_{bt}\}$.
Then we select $K_0$ largest-$\bar{\beta}_{kt}$ terminals for each active beam.
Accordingly, we decide $\alpha_{bt}$ and $\beta_{kt}$ for the current timeslot.
To update the residual demand $\bar{D}_k$, we solve the remaining of $\mathcal{P}_4(t)$ via Alg. \ref{alg:fp} with the decided integer solution.
At the end, with all determined $\alpha_{bt}$ and $\beta_{kt}$, we optimize $p_{kt}$ by solving $\mathcal{P}_1$ via Alg. \ref{alg:fp}.

\begin{algorithm}[t]
\caption{E-JPBT}
\label{alg:h}
\begin{algorithmic}[1]
\REQUIRE 
$\phi_{k}$. \\
\FOR{$t=1,\dots,T$}
\STATE Optimize $\bar{\alpha}_{bt}$ and $\bar{\beta}_{kt}$ by solving $\mathcal{P}_4(t)$ via Alg. \ref{alg:fp}. 
\STATE Select $j^*=\arg\max_{j}\{\sum_{b\in\mathcal{B}_j}\bar{\alpha}_{bt}\}$.
\STATE Set $\alpha_{bt}=1$, $\forall b\in\mathcal{B}_{j^*}$.
\STATE Select $K_0$ largest-$\bar{\beta}_{kt}$ terminals for each illuminated beam.
\STATE Set accordingly $\beta_{kt}=1$.
\STATE Optimize $p_{kt}$ by solving $\mathcal{P}_4(t)$ via Alg. \ref{alg:fp} with $\alpha_{bt}$ and $\beta_{kt}$.
\STATE Update $\bar{D}_{k}$ by \eqref{resid}.
\ENDFOR
\STATE Optimize $p_{kt}$ via Alg. \ref{alg:fp} with determined $\alpha_{bt}$ and $\beta_{kt}$.
\ENSURE
$\alpha_{bt}$, $\beta_{kt}$, and $p_{kt}$.
\end{algorithmic}
\end{algorithm}

The procedure of E-JPBT is summarized in Alg. \ref{alg:h}, where line 2 to line 6 are the process of determining integer variables for each timeslot.
In line 2, we solve the relaxed problem via Alg. \ref{alg:fp} and obtain the continuous solution, $\bar{\alpha}_{bt}$ and $\bar{\beta}_{kt}$.
The decisions of $\alpha_{bt}$ and $\beta_{kt}$ are described in line 4 and line 6, respectively.
With determined integer variables, power optimization is executed via Alg. \ref{alg:fp} to update $\bar{D}_k$.
At the end of E-JPBT, we optimize power with all the determined integer solution in line 7.
In conclusion, E-JPBT needs to apply Alg. \ref{alg:fp} for $2T+1$ times, and thus the complexity of E-JPBT is $\mathcal{O}((2T+1)\max\{N\psi\log(\frac{1}{\varepsilon}),\varrho^{-2}\})$.

\section{Performance Evaluation}

\subsection{Simulation Settings and Benchmarks}

\begin{table}[t]
\centering
\caption{Simulation parameters}
\captionsetup{font={footnotesize}} 
  \begin{tabular}{c|c}
  \hline
  {Parameter} & {Value}\\\hline
  Frequency, $f^{\mathrm{fr}}$ & 20 GHz (Ka band)\\
  Bandwidth, $W$ & 500 MHz\\
  Satellite location & $13^{\circ}$ E\\
  Satellite height & 35,786 km\\
  Beam radiation pattern & Provided by ESA \cite{esa}\\
  Power budget per beam, $P$ & 20 dBW \\
  Receive antenna gain, $G_{k}^{\textrm{rx}}$ & 42.1 dBi\\
  Noise power, $\sigma^2$ & -126.47 dBW\\
  Number of timeslots, $T$ & 256\\
  Number of beams, $B$ & 16, 37\\
  Maximum active beams, $B_0$ & 5\\
  Number of terminals per beam & 5 \\
  Minimum capacity, $R_{k}^{\min}$ & 5 Mbps\\
  Traffic demand, $D_{k}$ & 100 Mbps to 1.2Gbps\\
  Error ratio of imperfect SIC & $10^{-4}$ \\
  Maximum multiplexed terminals, $K_0$ & 3\\
  $N$, $N'$ & 20, 100 \\
  \hline
  \end{tabular}
\vspace{-0.2cm}
  \label{tab:para}
\end{table}

\begin{figure}[t]
\centering
\includegraphics[scale=0.5]{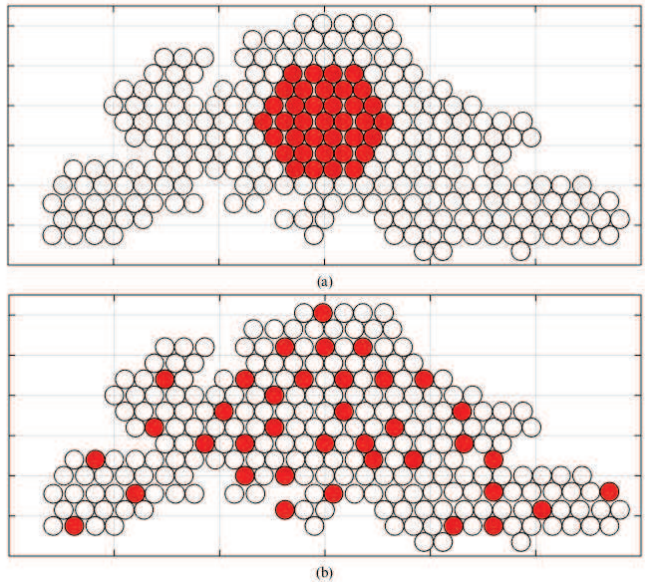}
\captionsetup{font={footnotesize}} 
\caption{Beam pattern covering Europe provided by ESA \cite{esa} and the adopted two scenarios, where BH is operated among the considered 37 beams (highlighted in red color).}
\label{fig:eu}
\end{figure}

In this section, we evaluate the performance of the proposed NOMA-BH scheme and the proposed algorithms in a multi-beam satellite system.
The parameter settings are summarized in Table \ref{tab:para} unless stated otherwise.
The beam radiation pattern is provided by European Space Agency (ESA) \cite{esa}, which is depicted in Fig. \ref{fig:eu}.
We consider two typical scenarios (highlighted in red color in Fig. \ref{fig:eu}) to evaluate the proposed schemes:
\begin{itemize}
\item Scenario 1 (Fig. \ref{fig:eu}(a)): We extract a set of adjacent beams from 245 beams, where the beam illumination design is carried out within a concentrated area.
\item Scenario 2 (Fig. \ref{fig:eu}(b)): We randomly select a set of non-adjacent beams from 245 beams, where BH is performed within a large area.
\end{itemize}
The results are averaged by 1000 instances.
In NOMA, we consider a practical issue in performance evaluation, i.e., residual interference in decoding terminals' signals due to imperfect SIC \cite{celik2018distributed}.  
Note that imperfect SIC is always considered in performance evaluation.
When calculating the offered-requested data rate gap, the units of capacity and demand are unified as Mbps.

We summarize the benchmark schemes as the following for different purposes of performance evaluation.
To investigate the benefits of combining BH and NOMA, we compare the proposed BH-NOMA schemes with the following standalone schemes either considering BH or NOMA (referring to Fig. \ref{fig:tdd} and Table II):
\begin{itemize}
\item BH-OMA (without NOMA): The BH-OMA problem can be formulated by simply restricting only one terminal at each timeslot, i.e., $K_0=1$ in $\mathcal{P}_0$, and then apply Alg. \ref{alg:sm} to obtain the optimized result.
\item 1c-NOMA (without BH): NOMA is adopted with 1-color frequency-reuse pattern (full-frequency reuse). 
All the beams keep illuminated without considering BH.
\item 2c-NOMA (without BH): NOMA is coordinated with 2-color frequency-reuse pattern where each color represents either vertical and horizontal polarization such that adjacent two beams can occupy orthogonal resources.
\item 4c-NOMA (without BH): NOMA is coordinated with 4-color frequency-reuse pattern. 
In the system, the frequency band is equally divided into two segments and each segment utilizes vertical and horizontal polarization. 
In this way, the adjacent four beams can occupy four different colors and the inter-beam interference can be reduced.
\end{itemize}

We also compare the performance achieved by the proposed algorithms with the following benchmarking schemes from the literature (referring to Fig. \ref{fig:md}):
\begin{itemize}
\item RA: We apply the resource allocation scheme proposed in \cite{ginesi2017joint} to determine the number of scheduled timeslots for each beam. 
Then Alg. \ref{alg:h} is applied to decide terminal-timeslot assignment and power allocation.
\item MaxSINR: An approach proposed in \cite{lei2011multibeam} and \cite{alegre2012offered} is adopted to determine the illuminated beams at each timeslot by selecting the beams with maximum SINR.
Then Alg. \ref{alg:h} is then adapted to optimize power allocation and terminal-timeslot assignment.
\item MinCCI: An efficient approach used in \cite{alegre2012offered} is applied to activate beams with the minimum inter-beam interference, then Alg. \ref{alg:h} is adopted analogously to MaxSINR.
\end{itemize}

For a fair comparison with other metrics for evaluating the offered-requested data rate matching, we adapt UBA to optimize the following widely-used objective functions (referring to Table III):
\begin{itemize}
\item Scheme 1: The objective is to max-min offered-capacity-to-requested-traffic ratio (OCTR), i.e., $\max\min_{k\in\mathcal{K}}\frac{R_k}{D_k}$, \cite{kibria2019precoded,wang2020noma,cocco2017radio} such that the worst capacity-demand mismatch effects among terminals can be mitigated.
\item Scheme 2:  The objective aims at minimizing the total unmet capacity of terminals, i.e., $\min\sum_{k\in\mathcal{K}}[D_k-R_k]^+$, \cite{Kisseleff2020radio,aravanis2015power,cocco2017radio}. 
\end{itemize}

\subsection{Benefits of Jointly Considering BH and NOMA}

\begin{figure}[t]
\centering
\includegraphics[scale=0.44]{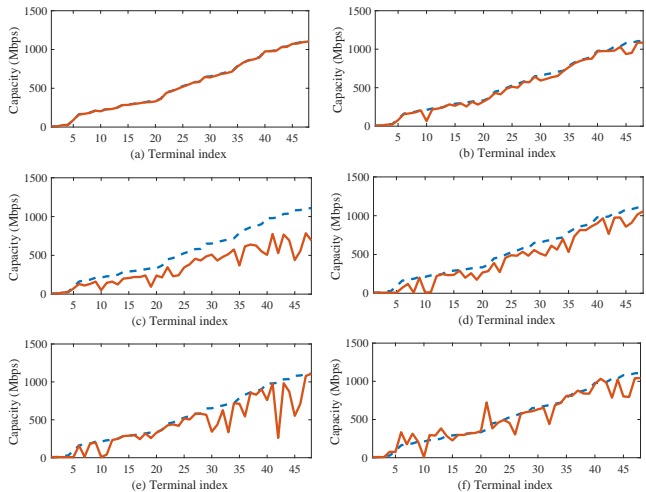}
\captionsetup{font={footnotesize}} 
\caption{An illustration of gaps between each user's data demand (blue dashed line) and the offered capacity (red solid line) obtained in:  (a) UBA, (b) E-JPBT, (c) BH-OMA, (d) 1c-NOMA, (e) 2c-NOMA, and (f) 4c-NOMA ($K=48$, $B=16$, $B_0=4$, and $K_0=2$).}
\label{fig:tdd}
\end{figure}

\begin{table}[t]
\centering
\caption{Power consumption (in Watts) of different schemes in Fig. \ref{fig:tdd}}
\captionsetup{font={footnotesize}} 
\centering
  \begin{tabular}{c|c|c|c|c|c}
  \hline
  UBA & E-JPBT & BH-OMA & 1c-NOMA & 2c-NOMA & 4c-NOMA \\\hline
  227.0 & 221.5 & 192.8 & 860.44 & 584.45 & 929.24 \\
  \hline
  \end{tabular}
\vspace{-0.2cm}
  \label{tab:pc}
\end{table}

In Fig. \ref{fig:tdd}, we discuss the benefits and evaluate the performance gains of combining BH and NOMA. 
The proposed BH-NOMA schemes, i.e., UBA and E-JPBT, are compared with the standalone schemes, either considering BH or NOMA.
The gaps are large in standalone NOMA or BH schemes, i.e., Fig. \ref{fig:tdd}(c) to Fig. \ref{fig:tdd}(f).
In contrast, by jointly optimizing BH and NOMA, the proposed BH-NOMA schemes in Fig. \ref{fig:tdd}(a) and Fig. \ref{fig:tdd}(b) significantly alleviate the mismatch effects, e.g., the objective value is reduced from $10^{5}-10^8$ in (Fig. \ref{fig:tdd}(c) -- Fig. \ref{fig:tdd}(f)) to $10^2-10^3$ (in Fig. \ref{fig:tdd}(a) and Fig. \ref{fig:tdd}(b)).
In Table \ref{tab:pc}, we further summarize the power consumption in these schemes.
BH-OMA consumes the least power, slightly lower than UBA and E-JPBT, but does not perform well in capacity-demand matching.
Compared to 1c-NOMA, 2c-NOMA, and 4c-NOMA, the proposed UBA and E-JPBT, by augmenting both power- and time-domain flexibilities, consume much less power and achieve good trade-offs between power saving and capacity-demand mismatch reduction.

\subsection{Performance in Bounding and Approaching the Optimum}

\begin{figure}[t]
\centering
\includegraphics[scale=0.8]{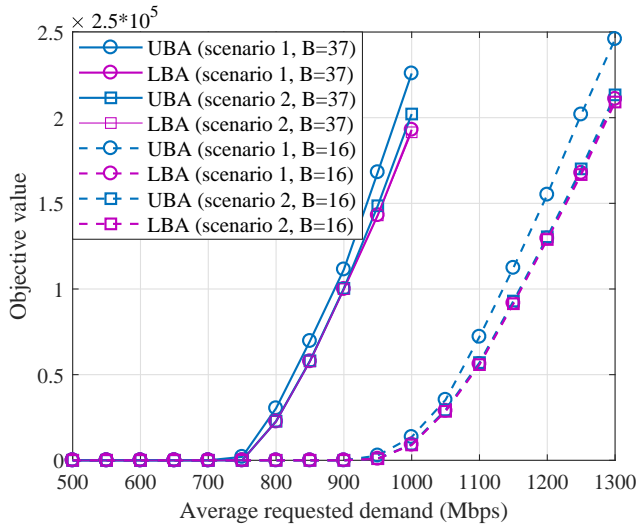}
\captionsetup{font={footnotesize}} 
\caption{The gap performance between upper bound and lower bound, where we set 5 terminals in each beam and $K_0=3$.}
\label{fig:ul}
\end{figure}

We evaluate the tightness of upper and lower bounds in Scenario 1 and Scenario 2 with 16 or 37 beams. 
From Fig. \ref{fig:ul}, we observe that the bounding gap increases as the average demand grows.
The proposed bounding scheme achieves near-zero gaps in Scenario 2, even for the cases with large demand.
This is because the inter-beam interference can maintain at a very low level when distant beams are activated.
When this small amount of inter-beam interference is intentionally neglected in LBA, it has limited impact and therefore keeps a tight lower bound for the optimum.
In contrast, when Scenario 1 is considered, a larger amount of inter-beam interference is removed in LBA, thus results in larger gaps, e.g., 14.9\% in 37 beams and 19.2\% in 16 beams.
The numerical results are consistent with the analysis in Remark 1.
By our design in LBA, less interference is neglected and a tighter lower bound can be obtained.  

\subsection{Performance Comparison between the Proposed Algorithms and Benchmarks}

\begin{figure}[t]
\centering
\includegraphics[scale=0.8]{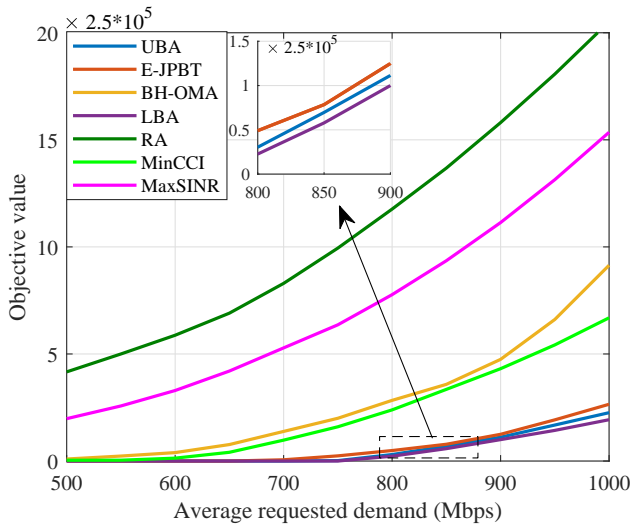}
\captionsetup{font={footnotesize}} 
\caption{The gap performance with respect to traffic demand among the proposed schemes and benchmarks. Here, the results are simulated in scenario 1 where there are 5 terminals in each beam with $K_0=3$.}
\label{fig:md}
\end{figure}

In Fig. \ref{fig:md}, we compare the performance of the proposed UBA, LBA, and E-JPBT with BH-OMA and benchmarking schemes from the literature.
We observe that the proposed schemes outperform all the four benchmarks in reducing the gap between requested and offered data rates.
Firstly, due to the higher spectral efficiency in NOMA, the proposed BH-NOMA schemes outperform BH-OMA, e.g., 80.8\% and 76.3\% improvement in UBA and E-JPBT.
Compared to RA, MaxSINR, and MinCCI, E-JPBT decreases the effect of offered-requested data mismatches by 93.2\%, 90.7\%, 70.4\%, respectively. 
The proposed BH-NOMA schemes can largely reduce the mismatch effects by joint optimization of BH and NOMA compared to GA, MaxSINR, and MinCCI. 
We also observe that E-JPBT can achieve the cost close to the upper bound with a gap of 18.95\% but with much less computational complexity compared to UBA.

\subsection{The Performance of the Proposed Schemes with Different Frequency-Reuse Patterns}

\begin{figure}[t]
\centering
\includegraphics[scale=0.8]{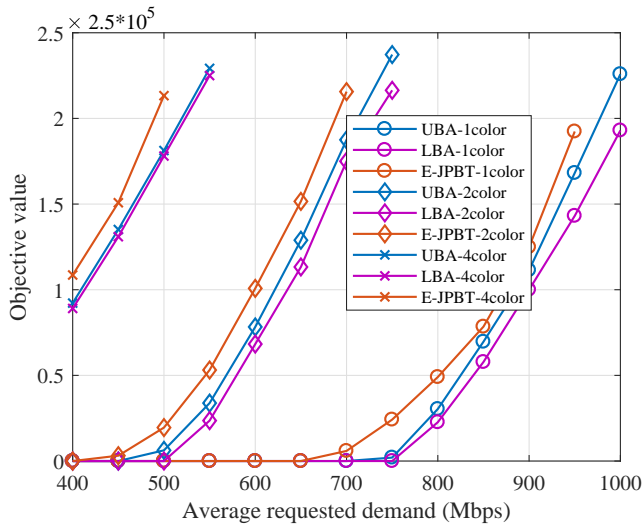}
\captionsetup{font={footnotesize}} 
\caption{The gap performance value with respect to traffic demand of the proposed BH-NOMA schemes in different scenarios with 1-color, 2-color, adn 4-color frequency-reuse patterns.}
\label{fig:color}
\end{figure}

In Fig. \ref{fig:color}, we evaluate the applicability of the proposed BH-NOMA schemes in the scenarios with the implementation of 1-color, 2-color, and 4-color frequency-reuse patterns.
The performance of the proposed schemes in all the three scenarios is promising. 
With higher spectral efficiency, the proposed schemes in 1-color scenario can perfectly match capacity to demand when the requested demand is no larger than 650 Mbps.
With less inter-beam interference, the average performance gaps between upper bound and lower bound in 2-color and 4-color scenarios are 11.23\% and 2.32\%, respectively, which are smaller than that in 1-color scenario.   
The result also verifies the conclusion in Remark 1.

\begin{table}[t]
\centering
\caption{The performance  comparison among different metrics}
\captionsetup{font={footnotesize}} 
\centering
  \begin{tabular}{l|c|c|c}
  \hline
  \diagbox{Metrics}{Schemes} & UBA & Scheme 1 & Scheme 2 \\\hline
  \tabincell{c}{Sum squared gaps \\ $\sum_{k\in\mathcal{K}}(R_{k}-D_{k})^2$ } & $1.57\times 10^{3}$ & $3.0\times 10^{4}$ & $1.69\times 10^{4}$ \\\hline
  The worst OCTR  & 0.93 & 0.96 & 0.88 \\\hline
  \tabincell{c}{Sum unmet capacity  \\ $\sum_{k\in\mathcal{K}}[D_k-R_k]^+$ } & 218.47 & 1013.181 & 215.30 \\
  \hline
  \end{tabular}
\vspace{-0.2cm}
  \label{tab:bnn}
\end{table}

\subsection{Comparison among Different Metrics for Evaluating the Offered-Requested Data Rate Mismatch}

Next, we compare the offered-requested data mismatch performance among different metrics.
In Table \ref{tab:bnn}, we summarize the performance, where each scheme is solved by its own objective, e.g., $1.57\times 10^3$  is obtained by UBA with the objective \eqref{po1} in the first row.
By summarizing the obtained solutions, the results for the other two metrics can be obtained, e.g., the worst OCTR for UBA in the second row.
As expected, all the three schemes perform the best with their own objectives, referring to the diagonal values, but can degrade sharply when measured with other metrics.
The proposed BH-NOMA scheme shows good adaptation and generalization capabilities among different metrics, which means that UBA achieves the best performance in reducing the sum squared gaps, and slightly losses around 1\% to 3\% performance in max-min OCTR and reducing unmet capacity than the other two schemes.

\subsection{Impact of imperfect SIC on BH-NOMA Performance}

\begin{figure}[t]
\centering
\includegraphics[scale=0.8]{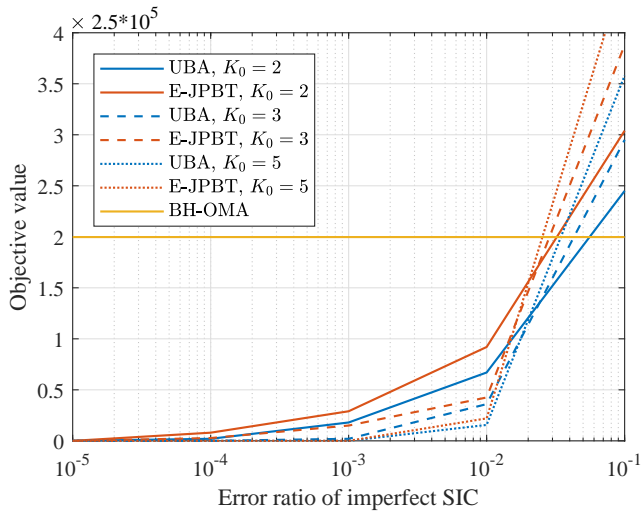}
\captionsetup{font={footnotesize}} 
\caption{The gap performance versus error ratio of imperfect SIC of the proposed NOMA schemes.}
\label{fig:eta}
\end{figure}

At last, we evaluate the impact of practical issues of NOMA on the performance of UBA and E-JPBT.
We introduce $0\leq\eta_k\leq 1$ to represent residual interference due to imperfect SIC \cite{celik2018distributed}.
Specifically, the intra-beam interference in \eqref{sinr} is rewritten as,
\begin{equation}
\sum\limits_{k{'}\in\mathcal{K}_b\setminus\{k\}\atop k'<k}|h_{bk}|^2p_{k't}+\sum\limits_{k{'}\in\mathcal{K}_b\setminus\{k\}\atop k'>k}|h_{bk}|^2p_{k't}\eta_k.
\end{equation} 
The result in Fig. \ref{fig:eta} shows the applicability of the proposed BH-NOMA schemes to imperfect-SIC scenarios.
We can observe that the performance increases slowly when the error ratio of imperfect SIC is small, e.g., from $10^{-5}$ to $10^{-2}$.
When the ratio increases more than $10^{-2}$, OMA might become a better choice.
Besides, when the error is large, e.g., $10^{-1}$, the mismatch effects in the case of $K_0=5$ is worse than those of $K_0=2$ and $K_0=3$. 
This is because the intra-beam interference caused by imperfect SIC increases with the number of co-channel terminals in the same beam.

\section{Conclusion}
In this paper, we have investigated joint resource optimization for the coexisted BH-NOMA systems.
A resource allocation problem has been formulated to minimize the gap between requested and offered data rates of terminals by jointly optimizing power allocation, beam scheduling, and terminal-timeslot assignment.
We have identified the NP-hardness of the problem and proposed an effective bounding scheme, UBA and LBA, to benchmark the optimality.
To reduce computational complexity, we have designed an efficient algorithm for joint optimization.
In the end, we have verified the benefits of combining BH and NOMA, and demonstrated the advantages of the proposed BH-NOMA schemes compared to different benchmarks.

\begin{appendices}

\begin{figure*}[t]
\begin{equation}
\label{dp}
\tilde{f}'(\zeta;p_{kt}^*)=\sum_{k\in\mathcal{K}_t^+}\frac{\frac{2(R_k-D_k)\gamma_{kt}}{\zeta(1+\gamma_{kt})}\left(\sum\limits_{k'\in\mathcal{K}_t^-}I_{k'kt}+\sigma^2\right)}{\sum\limits_{k'\in\mathcal{K}_t^+\setminus\{k\}}\zeta I_{k'kt}+\sum\limits_{k'\in\mathcal{K}_t^-}I_{k'kt}+\sigma^2}-\sum_{k\in\mathcal{K}_t^-}\frac{\frac{2(R_{k}-D_{k})\gamma_{kt}}{\zeta(1+\gamma_{kt})}\sum\limits_{k'\in\mathcal{K}_t^+}I_{k'kt}}{\sum\limits_{k'\in\mathcal{K}_t^+}\zeta I_{k'kt}+\sum\limits_{k'\in\mathcal{K}_t^-\setminus\{k\}}I_{k'kt}+\sigma^2}
\end{equation}
\hrulefill
\end{figure*}

\section{Proof of Lemma 1}
\label{lemma1}
\begin{proof}
We construct a polynomial-time reduction from three-dimensional matching (3DM) problem \cite{karp1972reducibility}, one of the typical NP-complete problems, to an instance of the decision-version problem of $\mathcal{P}_0$.
Consider three different sets $\mathcal{X}$, $\mathcal{Y}$, and $\mathcal{Z}$, where $|\mathcal{X}|=|\mathcal{Y}|=|\mathcal{Z}|$.
The 3DM problem is to check whether there exists a matching set ${\Theta}\subset\mathcal{X}\times\mathcal{Y}\times\mathcal{Z}$ such that $x_1\neq x_2$, $y_1\neq y_2$, and $z_1\neq z_2$ for any two different triplets $(x_1,y_1,z_1)$ and $(x_2,y_2,z_2)$ in $\Theta$.
If yes, $\Theta$ is called a 3DM.

Consider a special case with one terminal per beam, i.e., $K=B$.
In this case, we use terminals' indexes and beams' interchangeably.
The set of beams is divided into two subsets, $\mathcal{B}_1$ and $\mathcal{B}_2$, where $\mathcal{B}_1\cap\mathcal{B}_2=\varnothing$, $\mathcal{B}_1\cup\mathcal{B}_2=\mathcal{B}$, and $|\mathcal{B}_1|=|\mathcal{B}_2|=\frac{B}{2}$.
For any beam $b\in\mathcal{B}_i$, $\forall i=\{ 1,2\}$, the channel gains satisfy the following conditions,
\begin{equation}
|h_{b{'}b}|^2=
\begin{cases}
1+\epsilon, \textrm{ if $b'=b$,}\\
1+\frac{\epsilon}{2}, \textrm{ else if $b{'}\in\mathcal{B}_i$},\\
\epsilon, \textrm{ else if $b{'}\in\mathcal{B}_j, j\neq i$},
\end{cases}
\end{equation}
where $0<\epsilon\leq 2^{\frac{1}{B}}-1$.
We set the parameters as follows: 
$P=1$, $\sigma^2=\epsilon$, $B_0=2$, $T=\frac{B}{2}$, $\Omega=\varnothing$, $R^{\min}_{b}=1$, and $D_{b}\gg 1$.

First, we prove that the instance problem is feasible if the answer to the 3DM problem is yes.
We let $\mathcal{X}=\mathcal{T}$, $\mathcal{Y}=\mathcal{B}_1$, and $\mathcal{Z}=\mathcal{B}_2$.
For any two triplets $(t_1,b_1,b{'}_1)$ and $(t_2,b_2,b{'}_2)$ in $\Theta$, the following relationships hold: $t_1\neq t_2$, $b_1\neq b_2$, and $b{'}_1\neq b{'}_2$.
In this case, any two beams scheduled to the same timeslot are from different subsets.
If beam $b$ is illuminated at timeslot $t$, the rate of each beam is derived as $\log_2(1+\frac{|h_{bb}|^2P}{|h_{b{'}b}|^2P+\sigma^2})=\log_2 (1+\frac{1+\epsilon}{\epsilon+\epsilon})>1=R^{\min}_b$, which meets constraints \eqref{pc1} to \eqref{pc6}.
Thus the instance problem is feasible.

Next, we prove that if the instance problem is feasible, the answer to the 3DM problem is yes.
Since $T=\frac{B}{2}$ and $B_0=2$, all beams are scheduled only once.
If there exist two beams from the same subset scheduled to the same timeslot, then the rates for these two beams are $\log_2(1+\frac{|h_{bb}|^2P}{|h_{b{'}b}|^2P+\sigma^2})=\log_2(1+\frac{1+\epsilon}{1+\frac{\epsilon}{2}+\epsilon})<\log_2(1+1)=1=R^{\min}_b$, which violates the minimum-rate constraint in \eqref{pc5}.
To meet the constraints, the interference must be $\epsilon$, requiring that any two beams scheduled to the same timeslot are from different subsets.
Thus, the answer to the 3DM problem is yes.
In conclusion, the yes answer to the 3DM problem is the necessary and sufficient condition of the existence of a feasible solution of the instance problem.
As the 3DM problem is NP-complete, the Lemma follows.
\end{proof}

\section{Proof of Proposition 1}
\label{theorem2}
\begin{proof}
The proposition can be proven by raising the contradiction that there exist some terminals with $R_{k}>D_{k}$ at the optimum.
Define $\mathcal{K}_t$ as the set of the terminals scheduled to the $t$-th timeslot. 
We divide $\mathcal{K}_t$ into two subsets,
$\mathcal{K}_t^+$ and $\mathcal{K}_t^-$, containing terminals with $R_{k}> D_{k}$ and $R_{k}\leq D_{k}$, respectively.
Let $p_{kt}^*$ be the optimal power.
For presentation convenience, we denote $I_{k'kt}$ as the interference of terminal $k$ caused by $k'$ at timeslot $t$.
We apply $0<\zeta\leq 1$ to adjust the power of all the terminals in $\mathcal{K}_t^+$.
As $p_{kt}^*$ is optimal, $\zeta=1$ should be optimal.
For $k\in\mathcal{K}_t^+$, the SINR is expressed as,
\begin{equation}
\gamma_{kt}=\frac{|h_{bk}|^2\zeta p_{kt}^*}{\sum_{k'\in\mathcal{K}_t^+\setminus\{k\}}\zeta I_{k'kt}+\sum_{k'\in\mathcal{K}_t^-}I_{k'kt}+\sigma^2}.
\end{equation}
The SINR of terminals in $\mathcal{K}_t^-$ is,
\begin{equation}
\gamma_{kt}=\frac{|h_{bk}|^2 p_{kt}^*}{\sum_{k'\in\mathcal{K}_t^+}\zeta I_{k'kt}+\sum_{k'\in\mathcal{K}_t^-\setminus\{k\}}I_{k'kt}+\sigma^2}.
\end{equation}
Given $p_{kt}^*$, the objective can be viewed as the function of $\zeta$, say $\tilde{f}(\zeta;p_{kt}^*)$.
We present the derivative of $\tilde{f}(\zeta;p_{kt}^*)$ in \eqref{dp}, which is obviously larger than zero since $R_k>D_k$ for terminals in $\mathcal{K}_t^+$ and $R_k<D_k$ for terminals in $\mathcal{K}_t^-$.
Thus the objective can be smaller by letting $\zeta<1$ to reduce power of the terminals in $\mathcal{K}_t^+$, which contradicts the assumption of the optimality.
Thus the proposition.
\end{proof}

\end{appendices}





\bibliographystyle{IEEEtran}

\begin{thebibliography}{10}
\providecommand{\url}[1]{#1}
\csname url@samestyle\endcsname
\providecommand{\newblock}{\relax}
\providecommand{\bibinfo}[2]{#2}
\providecommand{\BIBentrySTDinterwordspacing}{\spaceskip=0pt\relax}
\providecommand{\BIBentryALTinterwordstretchfactor}{4}
\providecommand{\BIBentryALTinterwordspacing}{\spaceskip=\fontdimen2\font plus
\BIBentryALTinterwordstretchfactor\fontdimen3\font minus
  \fontdimen4\font\relax}
\providecommand{\BIBforeignlanguage}[2]{{%
\expandafter\ifx\csname l@#1\endcsname\relax
\typeout{** WARNING: IEEEtran.bst: No hyphenation pattern has been}%
\typeout{** loaded for the language `#1'. Using the pattern for}%
\typeout{** the default language instead.}%
\else
\language=\csname l@#1\endcsname
\fi
#2}}
\providecommand{\BIBdecl}{\relax}
\BIBdecl

\bibitem{wang2021joint}
A.~Wang, L.~Lei, E.~Lagunas, S.~Chatzinotas, A.~I. P{\'e}rez-Neira, and
  B.~Ottersten, ``Joint beam-hopping scheduling and power allocation in
  {NOMA}-assisted satellite systems,'' in \emph{2021 IEEE Wireless
  Communications and Networking Conference (WCNC)}.\hskip 1em plus 0.5em minus
  0.4em\relax IEEE, 2021, pp. 1--6.

\bibitem{maral2020satellite}
G.~Maral, M.~Bousquet, and Z.~Sun, \emph{Satellite communications systems:
  systems, techniques and technology}.\hskip 1em plus 0.5em minus 0.4em\relax
  John Wiley \& Sons, 2020.

\bibitem{kodheli2020satellite}
O.~Kodheli \emph{et~al.}, ``Satellite communications in the new space era: A
  survey and future challenges,'' \emph{IEEE Communications Surveys \&
  Tutorials}, vol.~23, no.~1, pp. 70--109, 2021.

\bibitem{cocco2017radio}
G.~Cocco, T.~De~Cola, M.~Angelone, Z.~Katona, and S.~Erl, ``Radio resource
  management optimization of flexible satellite payloads for{ DVB-S2}
  systems,'' \emph{IEEE Transactions on Broadcasting}, vol.~64, no.~2, pp.
  266--280, 2017.

\bibitem{angeletti2006beam}
P.~Angeletti, D.~Fernandez~Prim, and R.~Rinaldo, ``Beam hopping in multi-beam
  broadband satellite systems: System performance and payload architecture
  analysis,'' in \emph{24th AIAA International Communications Satellite Systems
  Conference}, 2006, p. 5376.

\bibitem{freedman2015beam}
A.~Freedman, D.~Rainish, and Y.~Gat, ``Beam hopping: how to make it possible,''
  in \emph{Proc. Ka and Broadband Communication Conference}, 2015.

\bibitem{kibria2019precoded}
M.~G. Kibria, E.~Lagunas, N.~Maturo, D.~Spano, and S.~Chatzinotas, ``Precoded
  cluster hopping in multi-beam high throughput satellite systems,'' in
  \emph{2019 IEEE Global Communications Conference (GLOBECOM)}.\hskip 1em plus
  0.5em minus 0.4em\relax IEEE, 2019, pp. 1--6.

\bibitem{lei2011multibeam}
J.~Lei and M.~A. Vazquez-Castro, ``Multibeam satellite frequency/time duality
  study and capacity optimization,'' \emph{Journal of Communications and
  Networks}, vol.~13, no.~5, pp. 472--480, 2011.

\bibitem{dvbs2x}
{ETSI EN 102 376–2}, ``Digital video broadcasting {(DVB)}; {DVB-S2X}
  implementation guidelines for the second generation system for broadcasting,
  interactive services, news gathering and other broadband satellite
  applications, part 2: {S}2-extensions ({DVB-S2X}),'' January 2021.

\bibitem{cao2019qos}
H.~Cao, Y.~Su, Y.~Zhou, and J.~Hu, ``{QoS} guaranteed load balancing in
  broadband multi-beam satellite networks,'' in \emph{ICC 2019-2019 IEEE
  International Conference on Communications (ICC)}.\hskip 1em plus 0.5em minus
  0.4em\relax IEEE, 2019, pp. 1--6.

\bibitem{zuo2018resource}
P.~Zuo, T.~Peng, W.~Linghu, and W.~Wang, ``Resource allocation for cognitive
  satellite communications downlink,'' \emph{IEEE Access}, vol.~6, pp.
  75\,192--75\,205, 2018.

\bibitem{9473538}
Y.~Wang, Y.~Chen, Y.~Qiao, H.~Luo, X.~Wang, R.~Li, and J.~Wang, ``Cooperative
  beam hopping for accurate positioning in ultra-dense {LEO} satellite
  networks,'' in \emph{2021 IEEE International Conference on Communications
  Workshops (ICC Workshops)}, 2021, pp. 1--6.

\bibitem{alegre2012offered}
R.~Alegre-Godoy, N.~Alagha, and M.~A. V{\'a}zquez-Castro, ``Offered capacity
  optimization mechanisms for multi-beam satellite systems,'' in \emph{2012
  IEEE International Conference on Communications (ICC)}.\hskip 1em plus 0.5em
  minus 0.4em\relax IEEE, 2012, pp. 3180--3184.

\bibitem{chen2021satellite}
L.~Chen, E.~Lagunas, S.~Chatzinotas, and B.~Ottersten, ``Satellite broadband
  capacity-on-demand: Dynamic beam illumination with selective precoding,'' in
  \emph{European Signal Processing Conference (EUSIPCO), Dublin, Ireland, Aug.
  2021}, 2021.

\bibitem{hu2020dynamic}
X.~Hu, Y.~Zhang, X.~Liao, Z.~Liu, W.~Wang, and F.~M. Ghannouchi, ``Dynamic beam
  hopping method based on multi-objective deep reinforcement learning for next
  generation satellite broadband systems,'' \emph{IEEE Transactions on
  Broadcasting}, vol.~66, no.~3, pp. 630--646, 2020.

\bibitem{lei2020beam}
L.~{Lei}, E.~{Lagunas}, Y.~{Yuan}, M.~G. {Kibria}, S.~{Chatzinotas}, and
  B.~{Ottersten}, ``Beam illumination pattern design in satellite networks:
  Learning and optimization for efficient beam hopping,'' \emph{IEEE Access},
  vol.~8, pp. 136\,655--136\,667, 2020.

\bibitem{yan2019application}
X.~Yan, K.~An, T.~Liang, G.~Zheng, Z.~Ding, S.~Chatzinotas, and Y.~Liu, ``The
  application of power-domain non-orthogonal multiple access in satellite
  communication networks,'' \emph{IEEE Access}, vol.~7, pp. 63\,531--63\,539,
  2019.

\bibitem{perez2019non}
A.~I. Perez-Neira, M.~Caus, and M.~A. Vazquez, ``Non-orthogonal transmission
  techniques for multibeam satellite systems,'' \emph{IEEE Communications
  Magazine}, vol.~57, no.~12, pp. 58--63, 2019.

\bibitem{wang2020noma}
A.~Wang, L.~Lei, E.~Lagunas, A.~I.~Pérez-Neira, S.~Chatzinotas, and
  B.~Ottersten, ``{NOMA}-enabled multi-beam satellite systems: Joint
  optimization to overcome offered-requested data mismatches,'' \emph{IEEE
  Transactions on Vehicular Technology}, vol.~70, no.~1, pp. 900--913, 2021.

\bibitem{chu2021robust}
J.~Chu, X.~Chen, C.~Zhong, and Z.~Zhang, ``Robust design for {NOMA}-based
  multibeam {LEO} satellite internet of things,'' \emph{IEEE Internet of Things
  Journal}, vol.~8, no.~3, pp. 1959--1970, 2021.

\bibitem{jiao2020network}
J.~Jiao, Y.~Sun, S.~Wu, Y.~Wang, and Q.~Zhang, ``Network utility maximization
  resource allocation for {NOMA} in satellite-based internet of things,''
  \emph{IEEE Internet of Things Journal}, vol.~7, no.~4, pp. 3230--3242, 2020.

\bibitem{Kisseleff2020radio}
S.~Kisseleff, E.~Lagunas, T.~S. Abdu, S.~Chatzinotas, and B.~Ottersten, ``Radio
  resource management techniques for multibeam satellite systems,'' \emph{IEEE
  Communications Letters}, pp. 1--1, 2020.

\bibitem{esa}
{ESA}, ``{SATellite Network of EXperts (SATNEX) IV},'' [Online],
  \url{https://satnex4.org/}.

\bibitem{panthi2017beam}
S.~Panthi, D.~Breynaert, C.~McLain, and J.~King, ``Beam hopping: {A} flexible
  satellite communication system for mobility,'' in \emph{35th AIAA
  International Communications Satellite Systems Conference}, 2017, p. 5413.

\bibitem{you2018resource}
L.~You, D.~Yuan, L.~Lei, S.~Sun, S.~Chatzinotas, and B.~Ottersten, ``Resource
  optimization with load coupling in multi-cell noma,'' \emph{IEEE Transactions
  on Wireless Communications}, vol.~17, no.~7, pp. 4735--4749, 2018.

\bibitem{boyd2004convex}
S.~Boyd, S.~P. Boyd, and L.~Vandenberghe, \emph{Convex optimization}.\hskip 1em
  plus 0.5em minus 0.4em\relax Cambridge university press, 2004.

\bibitem{liu2013complexity}
Y.-F. Liu and Y.-H. Dai, ``On the complexity of joint subcarrier and power
  allocation for multi-user {OFDMA} systems,'' \emph{IEEE transactions on
  Signal Processing}, vol.~62, no.~3, pp. 583--596, 2013.

\bibitem{hartmanis1982computers}
J.~Hartmanis, ``Computers and intractability: a guide to the theory of
  np-completeness (michael r. garey and david s. johnson),'' \emph{Siam
  Review}, vol.~24, no.~1, p.~90, 1982.

\bibitem{zappone2017globally}
A.~Zappone, E.~Bj{\"o}rnson, L.~Sanguinetti, and E.~Jorswieck, ``Globally
  optimal energy-efficient power control and receiver design in wireless
  networks,'' \emph{IEEE Transactions on Signal Processing}, vol.~65, no.~11,
  pp. 2844--2859, 2017.

\bibitem{shen2018fractional}
K.~Shen and W.~Yu, ``Fractional programming for communication systems—{P}art
  {I}: Power control and beamforming,'' \emph{IEEE Transactions on Signal
  Processing}, vol.~66, no.~10, pp. 2616--2630, 2018.

\bibitem{wang2012successive}
T.~Wang and L.~Vandendorpe, ``Successive convex approximation based methods for
  dynamic spectrum management,'' in \emph{2012 IEEE International Conference on
  Communications (ICC)}.\hskip 1em plus 0.5em minus 0.4em\relax IEEE, 2012, pp.
  4061--4065.

\bibitem{schurr2009polynomial}
S.~P. Schurr, D.~P. O'Leary, and A.~L. Tits, ``A polynomial-time interior-point
  method for conic optimization, with inexact barrier evaluations,'' \emph{SIAM
  Journal on Optimization}, vol.~20, no.~1, pp. 548--571, 2009.

\bibitem{zhao2016global}
R.~Zhao and J.~Fan, ``Global complexity bound of the levenberg--marquardt
  method,'' \emph{Optimization Methods and Software}, vol.~31, no.~4, pp.
  805--814, 2016.

\bibitem{lee2011mixed}
J.~Lee and S.~Leyffer, \emph{Mixed integer nonlinear programming}.\hskip 1em
  plus 0.5em minus 0.4em\relax Springer Science \& Business Media, 2011, vol.
  154.

\bibitem{di2016sub}
B.~Di, L.~Song, and Y.~Li, ``Sub-channel assignment, power allocation, and user
  scheduling for non-orthogonal multiple access networks,'' \emph{IEEE
  Transactions on Wireless Communications}, vol.~15, no.~11, pp. 7686--7698,
  2016.

\bibitem{celik2018distributed}
A.~Celik, M.-C. Tsai, R.~M. Radaydeh, F.~S. Al-Qahtani, and M.-S. Alouini,
  ``Distributed cluster formation and power-bandwidth allocation for imperfect
  {NOMA} in {DL-HetNets},'' \emph{IEEE Transactions on Communications},
  vol.~67, no.~2, pp. 1677--1692, 2018.

\bibitem{ginesi2017joint}
A.~Ginesi, E.~Re, and P.-D. Arapoglou, ``Joint beam hopping and precoding in
  hts systems,'' in \emph{International Conference on Wireless and Satellite
  Systems}.\hskip 1em plus 0.5em minus 0.4em\relax Springer, 2017, pp. 43--51.

\bibitem{aravanis2015power}
A.~I. Aravanis, B.~S. MR, P.-D. Arapoglou, G.~Danoy, P.~G. Cottis, and
  B.~Ottersten, ``Power allocation in multibeam satellite systems: A two-stage
  multi-objective optimization,'' \emph{IEEE Transactions on Wireless
  Communications}, vol.~14, no.~6, pp. 3171--3182, 2015.

\bibitem{karp1972reducibility}
R.~M. Karp, ``Reducibility among combinatorial problems,'' in \emph{Complexity
  of computer computations}, \hskip 1em plus 0.5em minus 0.4em\relax Springer,
  1972, pp. 85--103.

\end{thebibliography}


\end{document}